\documentclass[conference]{IEEEtran}

\newcommand{\e}{{\rm e}}
\usepackage{amsfonts}
\usepackage{color}
\usepackage{amsmath}
\usepackage{amsthm}
\usepackage{amssymb}
\usepackage{t1enc}
\usepackage{alltt}
\usepackage{epsfig}
\usepackage{mathrsfs}
\usepackage{graphicx,ifthen}
\usepackage{multirow}
\usepackage{citesort}
\usepackage[scaled]{helvet}

\newtheorem{lemma}{Lemma}
\newtheorem{proposition}{Proposition}

\makeatletter

\newcommand{\Rmnum}[1]{\expandafter\@slowromancap\romannumeral #1@}

\begin{document}
\title{Locally Best Invariant Test for Multiple Primary User Spectrum Sensing}

\author{\IEEEauthorblockN{Lu Wei, Prathapasinghe Dharmawansa and Olav Tirkkonen}
\IEEEauthorblockA{Department of Communications and Networking, \\
Aalto University\\
P. O. Box 13000, Aalto-00076, Finland \\
Email: \{lu.wei, olav.tirkkonen\}@aalto.fi, prathapakd@ieee.org}}

\maketitle

\begin{abstract}
We consider multi-antenna cooperative spectrum sensing in cognitive radio networks, when there may be multiple primary users. A noise-uncertainty-free detector that is optimal in the low signal to noise ratio regime is analyzed in such a scenario. Specifically, we derive the exact moments of the test statistics involved, which lead to simple and accurate analytical formulae for the false alarm probability and the decision threshold. Simulations are provided to examine the accuracy of the derived results, and to compare with other detectors in realistic sensing scenarios.
\end{abstract}
\


\section{Introduction}

In Cognitive Radio (CR) networks, dynamic spectrum access is implemented to mitigate spectrum scarcity. Namely, a secondary (unlicensed) user is allowed to utilize the spectrum resources when it does not cause intolerable interference to the primary (licensed) users. Spectrum sensing is the first key step towards this dynamic spectrum access scenario.

Prior work on cooperative spectrum sensing predominately employ the assumption of a single active primary user. Based on this assumption, several eigenvalue based sensing algorithms have been proposed recently~\cite{2008Yonghong,2010Taherpour,2009Lu,2008bZeng,2010Wang,2010Bianchi,2011Nadler,2011Lu}. These algorithms are non-parametric, i.e. they do not require information of the primary user, in contrast to e.g. feature detection. Moreover, they achieve optimality under different assumptions on the knowledge of the parameters. The assumption of a single primary user is made as the investigations in the literature have mainly focussed on CR networks, where the primary users are TV or DVB systems. In these systems the single active primary user assumption is, to some extent, justifiable. In addition, assuming a single primary user leads to analytically tractable problems.

The single primary user assumption may fail to reflect the situation in forthcoming CR networks, where the primary system could be a cellular network, and the existence of more than one primary user would be the prevailing condition. Using existing single primary user detection algorithms in such a scenario will induce performance loss. Despite the need to understand multiple primary user detection, the results in this direction are rather limited. A heuristic detection algorithm based on the ratio of the extreme eigenvalues is investigated in~\cite{2008Zeng,2009Federico}, but its detection performance turns out to be sub-optimal~\cite{2010Bianchi}. An optimal detection algorithm in the presence of multiple primary users, based on the spherical test, has been proposed in~\cite{2010Zhang} and subsequently studied in~\cite{2012Lu}. However, numerical evidence suggests that this detector does not perform particularly well when Signal-to-Noise Ratio (SNR) is relatively low~\cite{2012Lu}. Spectrum sensing in the low SNR regime is a practical and challenging issue in cooperative spectrum sensing. For example, recent FCC regulations require that the secondary devices must be able to detect signals with SNR as low as $-18$~dB~\cite{2010FCC}. To address this challenge, in this paper we consider a multiple primary user detector that is optimal in the low SNR regime. In particular, we investigate its detection performance by deriving a closed-form moment expression of the test statistics. Using the derived moments, approximations to the false alarm probability and the decision threshold are constructed. The derived approximations are easily computable and simulations show that they are accurate for the considered sensor sizes and number of samples.

The rest of this paper is organized as follows. In Section~\ref{sec:Formulation} we propose the optimal low SNR detector for multiple primary user sensing after outlining the signal model. Performance analysis of the proposed detection algorithm is addressed in Section~\ref{sec:Distributions}.
Section~\ref{sec:Simulations} presents numerical examples to verify the derived results and to study the detection performance in diverse scenarios. Finally in Section~\ref{sec:Conclusion} we conclude the main results of this paper.

\section{Problem Formulation}\label{sec:Formulation}

\subsection{Signal Model}

Consider the standard model for $K$-sensor cooperative detection in
the presence of $P$ primary users,
\begin{equation}\label{eq:model}
\mathbf{x}=\mathbf{Hs}+\sigma\mathbf{n}
\end{equation}
where $\mathbf{x}\in\mathbb{C}^{K}$ is the received data vector. The
$K$ sensors may be e.g. $K$ receive antennas in one secondary
terminal or $K$ secondary devices each with a single antenna, or any
combination of these.\footnote{This collaborative sensing scenario is more relevant when
  the $K$ sensors are in one device, since for multiple collaborating
  devices, accurate time synchronization between devices are needed
  and communications to the fusion center becomes an issue.}
 The $K\times P$ matrix
$\mathbf{H}=[\mathbf{h}_{1},\ldots,\mathbf{h}_{P}]$ represents the
channels between the $P$ primary users and the $K$ sensors. The
$P\times 1$ vector $\mathbf{s}=[s_{1},\ldots,s_{P}]'$ denotes zero mean transmitted signals
from the primary users. The $K\times 1$ vector $\sigma\mathbf{n}$ is the
complex Gaussian noise with zero mean and covariance matrix $\sigma^{2}\mathbf{I}_{K}$, where the scalar $\sigma^{2}$ is the noise power.

We collect $N$ i.i.d observations from model~(\ref{eq:model}) to a
$K\times N$ matrix $\mathbf{X} =
[\mathbf{x}_{1},\ldots,\mathbf{x}_{N}]$. The problem of interest is
to use the data matrix $\mathbf{X}$ to decide whether there are
primary users. For ease of analysis we make the following assumptions
\begin{enumerate}
\item The channel $\mathbf{H}$ is constant during sensing time.
\item The primary user's signal follows an i.i.d zero mean Gaussian distribution and is uncorrelated with the noise.
\end{enumerate}
In the absence of primary users, the
sample covariance matrix $\mathbf{R}=\mathbf{XX^{\dag}}$ follows an
uncorrelated (white) complex Wishart distribution
$\mathcal{W}_{K}\left(N,\mathbf{\Sigma}\right)$ with population
covariance matrix
\begin{equation}\label{eq:coH0}
\mathbf{\Sigma}:=\mathbb{E}[\mathbf{XX^{\dag}}]/N=\sigma^{2}\mathbf{I}_{K}.
\end{equation}
In the presence of primary users, by assumptions $1)$ and $2)$, the sample covariance matrix
$\mathbf{R}$ follows a correlated complex Wishart distribution. The
correlation is induced by the presence of the signals, and it is
characterized by the population covariance matrix
\begin{equation}\label{eq:coH1}
\mathbf{\Sigma}=\sigma^{2}\mathbf{I}_{K}+\sum_{i=1}^{P}\gamma_{i}\mathbf{h}_{i}\mathbf{h}^{\dag}_{i},
\end{equation}
where $\gamma_{i}:=\mathbb{E}[s_{i}s_{i}^{\dag}]$ defines the transmission power of the $i$-th primary
user. The received SNR of primary user $i$ across the $K$ sensors is $\text{SNR}_{i}:=\gamma_{i}||\mathbf{h}_{i}||^{2}/\sigma^{2}$. Finally, we denote the ordered eigenvalues of the sample covariance matrix $\mathbf{R}$ by $0\leq\lambda_K\leq\ldots\leq\lambda_{1}<\infty$.

\subsection{Test Statistics}

The differences between the population covariance matrices~(\ref{eq:coH0}) and~(\ref{eq:coH1}) can be explored to detect the primary users. This detection problem can be formulated as a binary hypothesis test, where hypothesis $\mathcal{H}_{0}$ denotes the absence of primary users and hypothesis $\mathcal{H}_{1}$
denotes the presence of primary users. Declaring wrongly $\mathcal{H}_{0}$, or declaring correctly $\mathcal{H}_{1}$, defines the false alarm probability $P_{\text{fa}}$, and the detection probability $P_{\text{d}}$, respectively. Since the sample covariance matrix $\mathbf{R}$ is a Wishart matrix, it is a sufficient statistics for the population covariance matrix $\mathbf{\Sigma}$~\cite{2003Anderson}. This leads to various test statistics as functions of $\mathbf{R}$ with different assumptions on the number of primary users $P$, and the knowledge of the noise power $\sigma^{2}$.

In the case of \emph{a single} primary user, $P=1$, the hypothesis
test can be expressed as
\begin{eqnarray}
\mathcal{H}_{0}&:& \mathbf{\Sigma}=\sigma^{2}\mathbf{I}_{K}\\
\mathcal{H}_{1}&:&
\mathbf{\Sigma}=\sigma^{2}\mathbf{I}_{K}+\gamma_{1}\mathbf{h}_{1}\mathbf{h}^{\dag}_{1}.
\end{eqnarray}
Assuming known noise power $\sigma^{2}$, the Largest Eigenvalue based
(LE) detection $T_{\text{LE}}:=\lambda_{1}$ is shown to be optimal
under the Generalized Likelihood Ratio Test (GLRT)
criterion~\cite{2010Taherpour}.
Here we concentrate on unknown noise
power. Then the optimal detector in the GLRT sense is the Scaled Largest
Eigenvalue based (SLE) detection
$T_{\text{SLE}}:=\lambda_{1}/\sum_{i=1}^{K}\lambda_{i}$. The SLE
detector is first proposed in the context of spectrum
sensing in~\cite{2008bZeng} and analyzed
in~\cite{2010Wang,2010Bianchi,2011Nadler,2011Lu}.

It can be verified that the matrix
$\sum_{i=1}^{P}\gamma_{i}\mathbf{h}_{i}\mathbf{h}^{\dag}_{i}$
in~(\ref{eq:coH1}) is positive definite, i.e.
$\sum_{i=1}^{P}\gamma_{i}\mathbf{h}_{i}\mathbf{h}^{\dag}_{i}\succ\mathbf{0}$.
Considering this fact, in the presence of possibly \emph{multiple}
primary users, when $P\geq 1$ but not known a priori, the hypothesis test is expressed as
\begin{eqnarray}
\mathcal{H}_{0}&:& \mathbf{\Sigma}=\sigma^{2}\mathbf{I}_{K}\\
\mathcal{H}_{1}&:& \mathbf{\Sigma}\succ\sigma^{2}\mathbf{I}_{K},
\end{eqnarray}
where the noise power $\sigma^{2}$ is assumed to be unknown. Essentially, we are testing a null hypothesis $\mathbf{\Sigma}=\sigma^{2}\mathbf{I}_{K}$ against all the other possible alternatives of $\mathbf{\Sigma}$, i.e. the hypothesis test is blind to $P$. In the statistics literature, this hypothesis test is known as the sphericity test. The corresponding optimal detector under the GLRT criterion is the so-called Spherical Test based detection $T_{\text{ST}}:=\prod_{i=1}^{K}\lambda_{i}\Big/\left(\sum_{i=1}^{K}\lambda_{i}/K\right)^{K}$. In the context of spectrum sensing, the ST detection is first proposed in~\cite{2010Zhang} and analyzed in~\cite{2012Lu}. Although in general the ST detector achieves good performance, it is not the best one in the low SNR regime. A test statistics that is optimal in detecting small deviations from $\mathcal{H}_{0}$ is the so-called John's detection
\begin{equation}\label{eq:TJ}
T_{\text{J}}:=\frac{\text{tr}(\mathbf{R}^{2})}{\big(\text{tr}(\mathbf{R})\big)^{2}}=\frac{\sum_{i=1}^{K}\lambda_{i}^{2}}{\left(\sum_{i=1}^{K}\lambda_{i}\right)^{2}},
\end{equation}
which is first considered by S. John~\cite{1971John}. A more rigorous derivation of the test statistics~(\ref{eq:TJ}) can be found in~\cite{1972Sugiura}, where the resulting test procedure is
\begin{equation}\label{eq:Hypo}
T_{\text{J}}\overset{\mathcal{H}_{1}}{\underset{\mathcal{H}_{0}}{\gtrless}}\zeta,
\end{equation}
$\zeta$ being a threshold. The optimality property of the $T_{\text{J}}$ detector in detecting small derivations is known as the locally best invariant property. Mathematically, it means that for every $\sigma^{2}$ and for every other test $T$ (say), there is a neighborhood of $\sigma^{2}\mathbf{I}_{K}$ such that $T_{\text{J}}$ achieves no worse performance than $T$ does~\cite{1972Sugiura}.

Besides the ST and John's detectors, another detector in the presence of multiple primary users is the Eigenvalue Ratio based (ER) detection $T_{\text{ER}}=\lambda_{1}/\lambda_{K}$~\cite{2008Zeng,2009Federico}. The ER detector is not constructed from any optimality considerations, thus its performance is substantially worse than those of the ST and John's detectors~\cite{2010Bianchi,2012Lu}. Finally, we note that no eigenvalue decomposition is needed in implementing John's detector as opposed to most of other eigenvalue based detectors.

\section{Performance Analysis}\label{sec:Distributions}

In this section we first derive an exact expression for the moment of $T_{\text{J}}$. Based on this result, we construct an approximation to the distribution of $T_{\text{J}}$, which leads to closed-form formulae for the false alarm probability and the decision threshold.

\subsection{Exact Moment Expression}

Under $\mathcal{H}_{0}$, by following the similar argument for the real Wishart case~\cite{1972John} it can be easily verified that the random variable $\left(\sum_{i=1}^{K}\lambda_{i}\right)^{2}$ is independent of the random variable of interest
\begin{equation}
T_{\text{J}}=\frac{\sum_{i=1}^{K}\lambda_{i}^{2}}{\left(\sum_{i=1}^{K}\lambda_{i}\right)^{2}}\in[1/K,1].
\end{equation}
By this independence, the $m$-th moment of $\sum_{i=1}^{K}\lambda_{i}^{2}$ equals
\begin{equation}
\mathbb{E}\Bigg[\left(\sum_{i=1}^{K}\lambda_{i}^{2}\right)^{m}\Bigg]=\mathbb{E}[T_{\text{J}}^{m}]\mathbb{E}\Bigg[\left(\sum_{i=1}^{K}\lambda_{i}\right)^{2m}\Bigg],
\end{equation}
and thus
\begin{equation}\label{eq:HH0}
\mathbb{E}[T_{\text{J}}^{m}]=\mathbb{E}\displaystyle\Bigg[\left(\sum_{i=1}^{K}\lambda_{i}^{2}\right)^{m}\Bigg]\Bigg/\mathbb{E}\displaystyle\Bigg[\left(\sum_{i=1}^{K}\lambda_{i}\right)^{2m}\Bigg].
\end{equation}

The random variable $2\,\text{tr}(\mathbf{R})=2\sum_{i=1}^{K}\lambda_{i}$ follows a Chi-square distribution with $2KN$ degrees of freedom. By using the moment expression for Chi-square distribution~\cite{2002Simon} (Eq. (2.35)), the $2m$-th moment of $\sum_{i=1}^{K}\lambda_{i}$ is obtained as
\begin{equation}\label{eq:mCHI}
\mathbb{E}\Bigg[\left(\sum_{i=1}^{K}\lambda_{i}\right)^{2m}\Bigg]=\frac{\Gamma(2m+KN)}{\Gamma(KN)}.
\end{equation}
The next step is to calculate the moment of $\sum_{i=1}^{K}\lambda_{i}^{2}$, which is given by the following result.
\begin{proposition}\label{th:mH0}
The $m$-th non-negative integer moment of the random variable $\sum_{i=1}^{K}\lambda_{i}^{2}$ equals
\begin{eqnarray}\label{eq:mH0}
\mathbb{E}\Bigg[\left(\sum_{i=1}^{K}\lambda_{i}^{2}\right)^{m}\Bigg]=\sum_{a_{1}+\cdots+a_{K}=m}\frac{m!C}{a_{1}!\cdots a_{K}!}\times\nonumber\\
\displaystyle\prod_{1\leq i<j \leq K}(2a_{j}-2a_{i}+j-i)\prod_{i=1}^{K}\Gamma(2a_{i}+N-K+i),&&
\end{eqnarray}
where the sum is over all the non-negative integer solutions of $a_{1}+\cdots+a_{K}=m$ and the constant $C=\left(\prod_{i=1}^{K}\Gamma(N-i+1)\Gamma(K-i+1)\right)^{-1}$.
\end{proposition}
The proof of Proposition~\ref{th:mH0} is in Appendix~\ref{ap:mH0}. Inserting~(\ref{eq:mH0}) and~(\ref{eq:mCHI}) into~(\ref{eq:HH0}), the $m$-th moment of random variable $T_{\text{J}}$, denoted by $\mathcal{M}_{m}$, equals
\begin{eqnarray}\label{eq:mTJ}
\mathcal{M}_{m}:=\frac{C\,\Gamma(KN)}{\Gamma(2m+KN)}\sum_{a_{1}+\cdots+a_{K}=m}\frac{m!}{a_{1}!\cdots a_{K}!}\times\nonumber\\
\displaystyle\prod_{1\leq i<j \leq K}(2a_{j}-2a_{i}+j-i)\prod_{i=1}^{K}\Gamma(2a_{i}+N-K+i).&&
\end{eqnarray}
The sum over the partition $a_{1}+\cdots+a_{K}=m$ can be implemented by normal sums as $\sum_{a_{1}=0}^{m}\sum_{a_{2}=0}^{m-a_{1}}\cdots\sum_{a_{K-1}=0}^{m-a_{1}-\cdots-a_{K-2}}$ with $a_{K}$ replaced by $m-\sum_{i=1}^{K-1}a_{i}$ in the summand. Note that for real Wishart matrix, up to the second, fourth and sixth moment of $T_{\text{J}}$ under $\mathcal{H}_{0}$ can be found in~\cite{1972John},~\cite{1976John} and~\cite{1993Boik}, respectively. To the best of our knowledge, the derived moment expression of $T_{\text{J}}$~(\ref{eq:mTJ}) for complex Wishart matrix, which is valid for arbitrary non-negative moment, is new.

\subsection{Moment Based Approximation}

It is a standard technique in statistics to approximate some unknown distribution by a known one having the same support and moments. Motivated by the results for real Wishart case~\cite{1993Boik}, in this work we choose a generalized Beta distribution with the same support as that of $T_{\text{J}}$ to approximate the distribution of $T_{\text{J}}$. Specifically, the linear transform $x=\frac{(K-1)z+1}{K}$ on a standard Beta density\footnote{$B(\alpha,\beta)=\Gamma(\alpha)\Gamma(\beta)/\Gamma(\alpha+\beta)$ defines the Beta function.} $z^{\alpha-1}(1-z)^{\beta-1}/B(\alpha,\beta)$, $z\in[0,1]$ leads to a generalized Beta density
\begin{equation}\label{eq:GBD}
C'\left(x-\frac{1}{K}\right)^{\alpha-1}(1-x)^{\beta-1},
\end{equation}
with the support $x\in[1/K,1]$ and the constant $C'=K^{\alpha+\beta-1}/\left(B(\alpha,\beta)(K-1)^{\alpha+\beta-1}\right)$. Since the $m$-th moment of a standard Beta random variable equals $\mathbb{E}[z^{m}]=(\alpha)_{m}/(\alpha+\beta)_{m}$, where $(\alpha)_{m}=\Gamma(\alpha+m)/\Gamma(\alpha)$ defines the Pochhammer symbol, the $m$-th moment of the generalized Beta random variable is obtained by binomial expansion as
\begin{eqnarray}
\mathbb{E}[x^{m}]&=&\mathbb{E}\left[\left(\frac{(K-1)z+1}{K}\right)^{m}\right]\\
&=&\frac{1}{K^{m}}\sum_{i=0}^{m}\binom{m}{i}(K-1)^{i}\mathbb{E}[z^{i}]\\
&=&\frac{1}{K^{m}}\sum_{i=0}^{m}\binom{m}{i}\frac{(K-1)^{i}(\alpha)_{i}}{(\alpha+\beta)_{i}}\label{eq:mGBD},
\end{eqnarray}
where $\binom{m}{i}$ denotes the binomial coefficient. In particular, the first two moments are
\begin{equation}
\frac{\alpha K+\beta}{(\alpha+\beta)K}, ~~~~\frac{(\alpha K+\beta)^2+\alpha K^{2}+\beta}{(\alpha+\beta)(\alpha+\beta+1)K^{2}},
\end{equation}
by matching them to the first two moments of $T_{\text{J}}$~(\ref{eq:mTJ}), the parameters $\alpha$ and $\beta$ of the generalized Beta density~(\ref{eq:GBD}) can be obtained
\begin{eqnarray}
\alpha &=& \frac{(K\mathcal{M}_{1}-1)(K\mathcal{M}_{1}-K\mathcal{M}_{2}+\mathcal{M}_{1}-1)}{(K-1)K(\mathcal{M}_{2}-\mathcal{M}_{1}^{2})}, \\
\beta &=& \frac{(\mathcal{M}_{1}-1)(K\mathcal{M}_{1}-K\mathcal{M}_{2}+\mathcal{M}_{1}-1)}{(K-1)(\mathcal{M}_{1}^{2}-\mathcal{M}_{2})}.
\end{eqnarray}
As a result, the two-moment-based approximation to the CDF of
$T_{\text{J}}$ under $\mathcal{H}_{0}$ is
\begin{eqnarray}
F_{\text{J}}(y) &\approx& C'\int_{1/K}^{y}\left(x-\frac{1}{K}\right)^{\alpha-1}(1-x)^{\beta-1}\mathrm{d}x \\
  &=& 1-\frac{B\left(\frac{K(1-y)}{K-1};\beta,\alpha\right)}{B(\alpha,\beta)},
\end{eqnarray}
where $y\in[1/K,\infty)$ and $B(x;a,b)=\int_{0}^{x}z^{a-1}(1-z)^{b-1}\mathrm{d}z$ is the lower incomplete Beta function.

By~(\ref{eq:Hypo}), for a given threshold $\zeta$ the two-moment-based approximation to the false alarm probability equals
\begin{equation}\label{eq:Pfa}
P_{\text{fa}}(\zeta)=1-F_{\text{J}}(\zeta)\approx\frac{B\left(\frac{K(1-\zeta)}{K-1};\beta,\alpha\right)}{B(\alpha,\beta)}.
\end{equation}
To implement the proposed spectrum sensing algorithm, a decision threshold needs to be determined for a given detection requirement in a computationally affordable manner. Using the derived approximation to $P_{\text{fa}}$, an approximative decision threshold can be obtained by numerically inverting~(\ref{eq:Pfa}). Note that both $\alpha$ and $\beta$ are elementary functions of the sensor size $K$ and sample size $N$ through~(\ref{eq:mH0}). Moreover, if we further approximate $\alpha$ and $\beta$ to their respective nearest integers, (\ref{eq:Pfa}) reduces to a finite sum of polynomials in $\zeta$. Thus the computational complexity of threshold calculation becomes quite affordable for on-line implementations.

Here we note that the proposed two-moment-based Beta approximation
corresponds to the simplest form of a general Jacobi polynomials based
approximation. In the general framework, up to any $n$-th degree of
Jacobi polynomials matching the corresponding $n$ moments of
$T_{\text{J}}$ would be used. Since the random variable $T_{\text{J}}$ is of a finite support, the Jacobi polynomials expansion for the distribution of $T_{\text{J}}$ is exact according to Weierstrass approximation theorem~\cite{1961Hochstadt}. Namely, when $n$ goes to infinity the Jacobi polynomials based approximation represents the exact distribution of $T_{\text{J}}$. In practise, the choice of $n$ reflects a trade-off between the approximation accuracy and the implementation complexity. In light of the good accuracy as shown in the next section, we consider $n=2$ in this paper. The general $n$-moment-based approximation, including the error analysis, can be easily obtained by following the procedures in~\cite{1993Boik,2006Ha}.

\section{Numerical Results}\label{sec:Simulations}

In this section we first investigate the accuracy of the derived approximative false alarm probability by simulations. Then we compare the performance of John's detector with those of several detectors in realistic scenarios. The considered values of the parameters $K$ and $N$ in this section reflect practical spectrum sensing situations. The sample size $N$ can be as large as a couple of hundred whereas the number of sensors $K$ is at most eight due to physical constraints of the device size.

\begin{figure}[t!]
\centering
\includegraphics[width=3.5in]{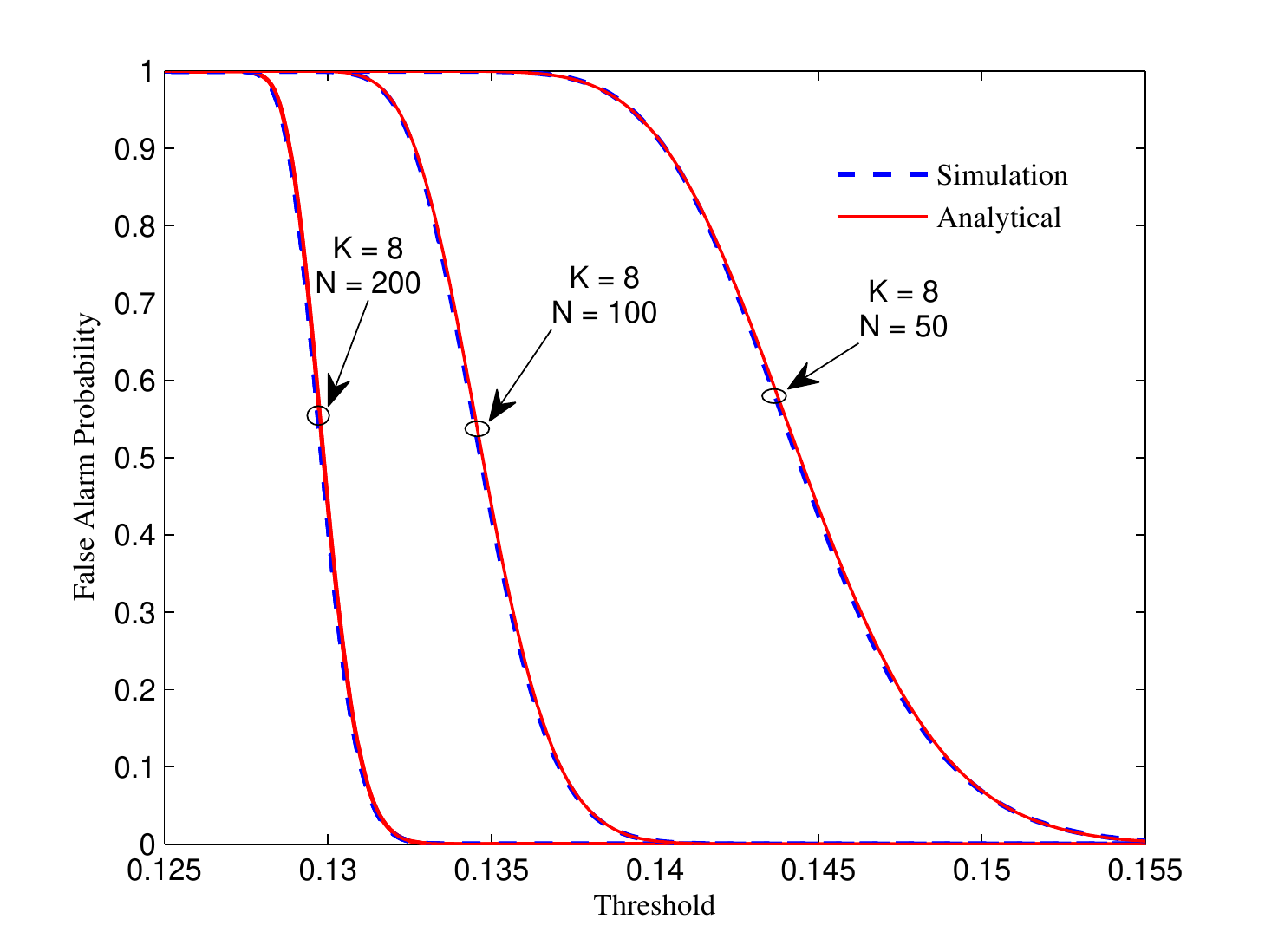}
\caption{False alarm probability: analytical versus simulations. For $K=8$, $N=50$, $100$ and $200$, the average approximation error on false alarm probability is respectively $5.94\times10^{-8}$, $5.03\times10^{-8}$ and $5.07\times10^{-8}$.}\label{fig:Pfa}
\end{figure}

\subsection{Accuracy of the Approximative False Alarm Probability}\label{subsec:Pfa}
In Figure~\ref{fig:Pfa} we plot the approximative $P_{\text{fa}}$ using~(\ref{eq:Pfa}) and the simulated $P_{\text{fa}}$ as a function of the threshold. To quantitatively show the approximation accuracy, we also calculate the average approximation error\footnote{Defined as $\left(\sum_{i=1}^{n}|P_{\text{fa}}(\zeta_{i})-\tilde{P}_{\text{fa}}(\zeta_{i})|\right)/n$, where $\tilde{P}_{\text{fa}}$ denotes the approximative false alarm probability, i.e. the RHS of~(\ref{eq:Pfa}), and $n$ is the sampling size. In Figure~\ref{fig:Pfa} we assume uniform sampling in $\zeta\in[0.125,0.3]$ with $n=10^{7}$.} of the proposed $P_{\text{fa}}$ with respect to the exact $P_{\text{fa}}$ as resulting from simulations. The results, summarized in the caption of Figure~\ref{fig:Pfa}, show that the derived analytical $P_{\text{fa}}$ matches the simulations well.

\subsection{Detection Performance}
We compare the detection performance of John's detector with those of other known detectors by means of ROC curve. Since a ROC curve shows the achieved detection probability as a function of the false alarm probability, it reflects the overall detection performance for a given detector. Our focus here is detection in the presence of multiple primary users, thus we consider for comparison the ST detector. In addition the SLE detector, which is optimal for single primary user detection, is considered for comparison as well. Comparisons with the non-optimal ER detector and detectors that are sensitive to noise uncertainty~\cite{2008Tandra}, such as the LE detector and the energy detector~\cite{2003Digham}, are excluded in this paper. For results in this direction, the readers are referred to~\cite{2010Bianchi,2011Nadler,2012Lu}.
\begin{figure}[t!]
\centering
\includegraphics[width=3.5in]{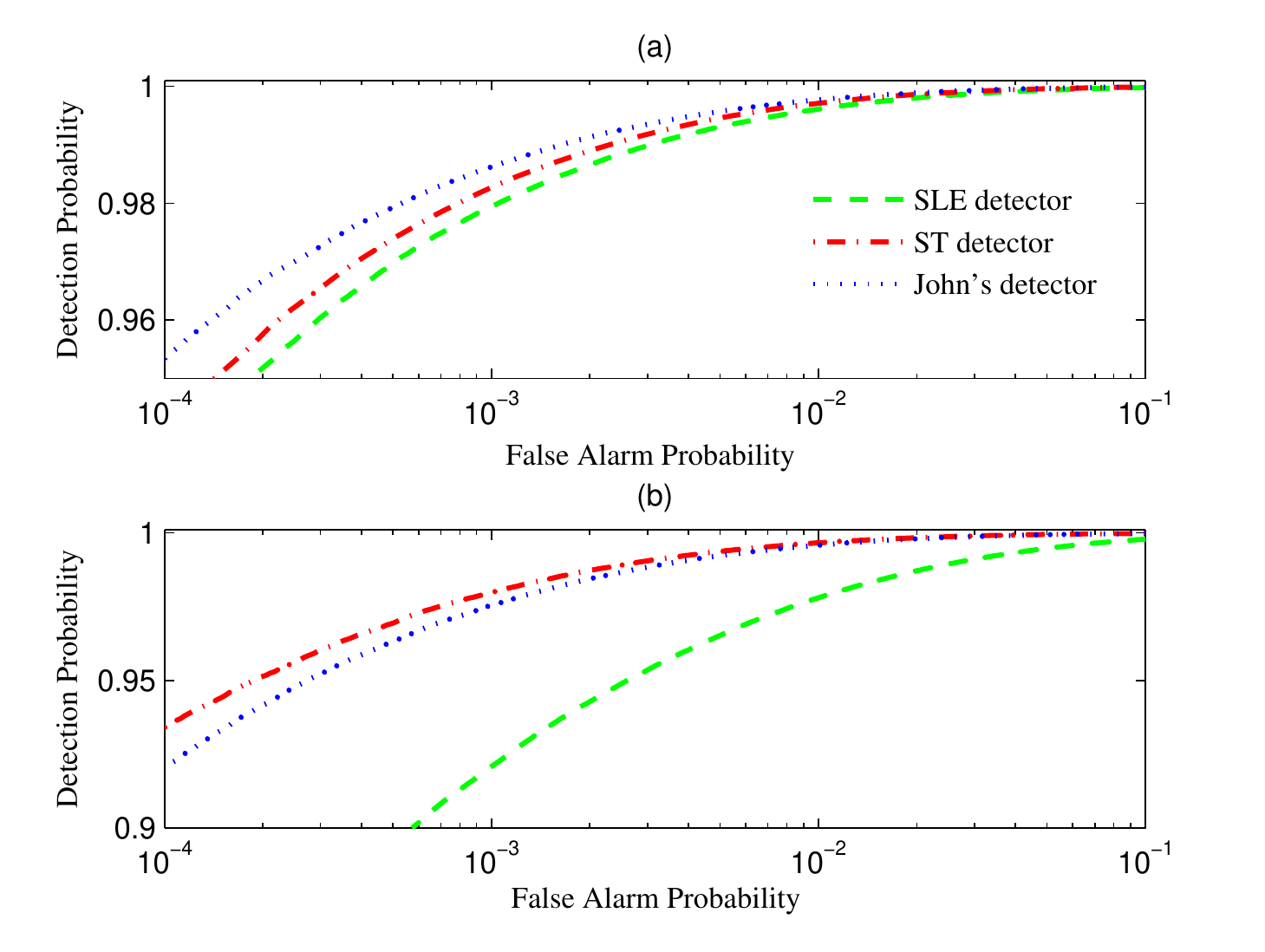}
\caption{ROCs for $P=3$. Subplot (a): $\text{SNR}_{1}=-6$~dB, $\text{SNR}_{2}=-5$~dB, $\text{SNR}_{3}=-4$~dB with $K=4$, $N=400$; subplot (b): $\text{SNR}_{1}=1$~dB, $\text{SNR}_{2}=2$~dB, $\text{SNR}_{3}=3$~dB with $K=4$, $N=50$.}\label{fig:ROC}
\end{figure}

In Figure~\ref{fig:ROC} we consider a scenario of three simultaneously transmitting primary users with relatively low SNRs ($\text{SNR}_{1}=-6$~dB, $\text{SNR}_{2}=-5$~dB, $\text{SNR}_{3}=-4$~dB using $K=4$, $N=400$) in subplot (a) and relatively high SNRs ($\text{SNR}_{1}=1$~dB, $\text{SNR}_{2}=2$~dB, $\text{SNR}_{3}=3$~dB using $K=4$, $N=50$) in subplot (b). Without loss of generality, we assume unit powers for the zero mean Gaussian signal and noise. The entries of the channel matrix $\mathbf{H}$, which are fixed during sensing, are independently drawn from a standard complex Gaussian distribution. The same channel matrix is used in both subplots. The channel vector for each primary user is normalized as $\mathbf{u}_{i}=\mathbf{h}_{i}/||\mathbf{h}_{i}||$. As a result, the population covariance matrix $\mathbf{\Sigma}$ now equals $\mathbf{\Sigma}=\mathbf{I}_{K}+\sum_{i=1}^{P}\text{SNR}_{i}\mathbf{u}_{i}\mathbf{u}^{\dag}_{i}$. For the specific channel realizations considered in Figure~\ref{fig:ROC}, the eigenvalues\footnote{For the considered detectors, the test statistics depend on $\mathbf{\Sigma}$ only through the eigenvalues of $\mathbf{\Sigma}$.} of $\mathbf{\Sigma}$ in subplot (a) are $[1.6225, 1.2217, 1.1213, 1]$ in subplot (b) are $[4.0417, 2.2375, 1.56,1]$. From Figure~\ref{fig:ROC} (a) we observe that John's detector achieves the best detection performance in the low SNR case considered. However, when the SNRs increase we see from Figure~\ref{fig:ROC} (b) that the ST detector performs better than John's detector. In both subplots, it is seen that the ST and John's detectors outperform the SLE detector. This is intuitively clear since the SLE detector is optimized for single primary user detection. Moreover, we see that as the SNRs increase the performance gap between the SLE detector and the multiple primary user detectors becomes larger, as expected.

\section{Conclusion}\label{sec:Conclusion}

In this paper, we investigated the sensing performance of John's
detector, which a candidate detector in the presence of multiple
primary users. John's detector is optimal in detecting small
deviations of the covariance matrix  from a matrix proportional to identity. Analytical formulae have been derived for the false alarm probability and decision threshold of John's detector. The derived results are simple to calculate and yield an almost exact fit to simulations. From the simulation setting considered, performance gain over several detection algorithms is observed in the low SNR regime.

Characterizing $T_{\text{J}}$ distribution under $\mathcal{H}_{1}$, which leads to analytical results for detection probability, is the work in progress.

\section*{Acknowledgment}
This work is partially supported by the Academy of Finland.

\appendices

\section{The non-negative integer moment of $\sum_{i=1}^{K}\lambda_{i}^{2}$}\label{ap:mH0}
Under $\mathcal{H}_{0}$, the joint density of the unordered eigenvalues $\lambda_{i}\in[0,\infty)$ for the sample covariance matrix $\mathbf{R}$ reads~\cite{1964James}
\begin{equation}
\Phi(\lambda_{1},\ldots,\lambda_{K}):=\frac{C}{K!}\left\vert\Delta(\mathbf{\lambda})\right\vert^{2}\displaystyle\prod_{i=1}^{K}\lambda_{i}^{N-K}\e^{-\lambda_{i}},
\end{equation}
where $\left\vert\Delta(\mathbf{\lambda})\right\vert$ defines the determinant of Vandermonde matrix with $i,j$-th entry $(\Delta(\mathbf{\lambda}))_{i,j}=\lambda_{i}^{j-1}$, $i,j=1,\ldots,K$ and the constant $C=\left(\prod_{i=1}^{K}\Gamma(N-i+1)\Gamma(K-i+1)\right)^{-1}$.

Before we prove Proposition~\ref{th:mH0}, we need the following two lemmas.
\begin{lemma}\label{th:lemINT}
The average value of the function $\prod_{i=1}^{K}\lambda_{i}^{2a_{i}}$, where $a_{i}$s are non-negative integers, equals
\begin{eqnarray}\label{eq:GSI}
\int_{[0,\infty)^{K}}\left(\prod_{i=1}^{K}\lambda_{i}^{2a_{i}}\right)\Phi(\lambda_{1},\ldots,\lambda_{K})\mathrm{d}\lambda_{1}\cdots\mathrm{d}\lambda_{K}=\nonumber\\
\frac{C}{K!}\sum_{\nu}\left\vert\Gamma(2a_{\nu_{i}}+N-K+i+j-1)\right\vert_{i,j=1,\ldots,K},
\end{eqnarray}
where $\nu=\nu_{1},\ldots,\nu_{K}$ defines a permutation of the integers $1,\ldots,K$ and the sum is over all the $K!$ permutations.
\end{lemma}

\begin{IEEEproof}
Using the fact that a product of determinants equals the determinant of the matrix product and the fact that the determinant remains unchanged under transpose operation, we have
\begin{equation}
|\Delta(\mathbf{\lambda})|^{2}=|\Delta(\mathbf{\lambda})\Delta'(\mathbf{\lambda})|=\left\vert\sum_{l=1}^{K}\lambda_{l}^{i+j-2}\right\vert_{i,j=1,\ldots,K}.\label{eq:Hankel}
\end{equation}
By invoking the multi-linearity property of determinants, the Hankel
determinant~(\ref{eq:Hankel})  above can be written as a sum of two determinants, where the first rows are $[1,\cdots,\lambda_{1}^{K-1}]$ and $[K-1,\cdots,\sum_{i=2}^{K-1}\lambda_{i}^{K-1}]$ with the respect second to the last rows remain unchanged. By repeated use of the multi-linearity property, (\ref{eq:Hankel}) can be written as sum of $K^{K}$ determinants, out of which $K!$ determinants with non-identical $\lambda$ index in rows give non-zero contribution, namely,
\begin{equation}\label{eq:sumdet}
\left\vert\sum_{l=1}^{K}\lambda_{l}^{i+j-2}\right\vert_{i,j=1,\ldots,K}=\sum_{\nu}\left\vert\lambda_{\nu_{i}}^{i+j-2}\right\vert_{i,j=1,\ldots,K},
\end{equation}
where $\nu=\nu_{1},\ldots,\nu_{K}$ defines a permutation of the integers $1,\ldots,K$ and the sum is over all the $K!$ permutations. Inserting (\ref{eq:sumdet}) into LHS of (\ref{eq:GSI}) and disregarding the constant $C/K!$ which will be re-installed, we have
\begin{eqnarray*}
&&\!\!\!\int_{[0,\infty)^{K}}\sum_{\nu}\left\vert\lambda_{\nu_{i}}^{i+j-2}\right\vert_{i,j=1,\ldots,K}\displaystyle\prod_{i=1}^{K}\lambda_{i}^{2a_{i}+N-K}\e^{-\lambda_{i}}\mathrm{d}\lambda_{i}\\ &=&\!\!\!\sum_{\nu}\int_{[0,\infty)^{K}}\left\vert\lambda_{\nu_{i}}^{2a_{\nu_{i}}+N-K+i+j-2}\e^{-\lambda_{\nu_{i}}}\right\vert_{i,j=1,\ldots,K}\displaystyle\prod_{i=1}^{K}\mathrm{d}\lambda_{i}\\
&=&\!\!\!\sum_{\nu}\left\vert\Gamma(2a_{\nu_{i}}+N-K+i+j-1)\right\vert_{i,j=1,\ldots,K},
\end{eqnarray*}
where in the first equality we multipled each $\lambda_{i}^{2a_{i}+N-K}\e^{-\lambda_{i}}$ with the row of $\left\vert\lambda_{\nu_{i}}^{i+j-2}\right\vert_{i,j=1,\ldots,K}$ having the same $\lambda$ index and the second equality is achieved by first expanding the determinant using Leibniz formula, integrating term-wise and rewriting the integration results as a determinant. This completes the proof.
\end{IEEEproof}
Note that Lemma~\ref{th:lemINT} can be considered as an extension to the Selberg type integral considered in~\cite{Mehta} (Eq. (17.6.5) and Eq. (17.8.1)).

\begin{lemma}\label{th:lemDET}
The following determinant can be simplified to
\begin{equation}
\left\vert\Gamma(b_{i}+j-1)\right\vert_{i,j=1,\ldots,K}=\displaystyle\prod_{1\leq i<j \leq K}(b_{j}-b_{i})\prod_{i=1}^{K}\Gamma(b_{i}),
\end{equation}
where $b_{i}$ is a positive integer.
\end{lemma}

\begin{IEEEproof}
We first realize that from each row the term $\Gamma(b_{i})$ can be factored out, namely,
\begin{equation}
\left\vert\Gamma(b_{i}+j-1)\right\vert_{i,j=1,\ldots,K}=\left\vert(b_{i})_{j-1}\right\vert_{i,j=1,\ldots,K}\displaystyle\prod_{i=1}^{K}\Gamma(b_{i}),
\end{equation}
where $(b)_{j}=\prod_{k=0}^{j-1}(b+k)$. By extracting from the $i$-th column a suitable linear combination of previous $i-1$ columns, the determinant
\begin{equation}
\left\vert(b_{i})_{j-1}\right\vert_{i,j=1,\ldots,K}=\left\vert b_{i}^{j-1}\right\vert_{i,j=1,\ldots,K},
\end{equation}
which is a Vandermonde determinant $\prod_{1\leq i<j \leq K}(b_{j}-b_{i})$. This completes the proof.
\end{IEEEproof}

We are now in a position to prove Proposition~\ref{th:mH0}. By using the multinomial expansion
\begin{equation}
\left(\sum_{i=1}^{K}\lambda_{i}^{2}\right)^{m}=\sum_{a_{1}+\cdots+a_{K}=m}\frac{m!}{a_{1}!\cdots a_{K}!}\prod_{i=1}^{K}\lambda_{i}^{2a_{i}}
\end{equation}
and Lemma~\ref{th:lemINT}, we have
\begin{eqnarray}\label{eq:GmH0}
\mathbb{E}\Bigg[\left(\sum_{i=1}^{K}\lambda_{i}^{2}\right)^{m}\Bigg]=\sum_{a_{1}+\cdots+a_{K}=m}\frac{m!}{a_{1}!\cdots a_{K}!}\frac{C}{K!}\times\nonumber\\
\sum_{\nu}\left\vert\Gamma(2a_{\nu_{i}}+N-K+i+j-1)\right\vert_{i,j=1,\ldots,K}.&&
\end{eqnarray}
For any given permutation $\nu$, it is observed that the sum over $a_{1}+\cdots+a_{K}=m$ is symmetric in the sense that one can arbitrarily permute the index of $a$ without changing the value of this sum, in particular the following permutation of $a$ holds,
\begin{eqnarray*}
&&\!\!\!\!\!\!\!\!\sum_{a_{1}+\cdots+a_{K}=m}\!\!\!\frac{m!\left\vert\Gamma(2a_{\nu_{i}}+N-K+i+j-1)\right\vert_{i,j=1,\ldots,K}}{a_{1}!\cdots a_{K}!}\\
&=&\!\!\!\!\!\!\!\!\sum_{a_{1}+\cdots+a_{K}=m}\!\!\!\frac{m!\left\vert\Gamma(2a_{i}+N-K+i+j-1)\right\vert_{i,j=1,\ldots,K}}{a_{1}!\cdots a_{K}!}.
\end{eqnarray*}
Since the number of possible permutations is $K!$, (\ref{eq:GmH0}) now equals
\begin{equation}
\sum_{a_{1}+\cdots+a_{K}=m}\frac{m!C\left\vert\Gamma(2a_{i}+N-K+i+j-1)\right\vert_{i,j=1,\ldots,K}}{a_{1}!\cdots a_{K}!}.
\end{equation}
Using Lemma~\ref{th:lemDET} with $b_{i}=2a_{i}+N-K+i$ completes the proof of Proposition~\ref{th:mH0}.

\ifCLASSOPTIONcaptionsoff
\newpage
\fi

\IEEEtriggeratref{0}

\end{document}